\documentclass[12pt,thmsa]{article}
\usepackage{amsfonts}
\usepackage{graphicx}
\usepackage{epsfig}
\usepackage{graphics}

\makeatletter
\newcommand{\row}[1]{\mathord{\buildrel{\lower3pt\hbox{$\scriptscriptstyle\rightarrow$}}\over #1}}

\newcommand{\dyadic}[1]{\mathord{\dyadic@rrow{#1}}}
\newcommand{\dyadic@rrow}[1]{
\begin{picture}(12,12)(-1,0)
\put(-3,12){\makebox(0,0)[t]{$\scriptscriptstyle\downarrow$}}
\put(-3,13){\makebox(0,0)[l]{$\scriptscriptstyle\longrightarrow$}}
\put(5,0){\makebox(0,0)[b]{$#1$}}
\end{picture}
}
\newcommand{\bra}[1]{\bigl\langle #1 \bigr|}
\newcommand{\ket}[1]{\bigl| #1 \bigr\rangle}

\begin{document}

\date{\today}

\begin{center}
{ Dynamics of multi-qubit systems in noisy environment}\\[0pt]
\vspace{0.5cm}  N. Metwally$^{1,2}$ and A. Almannaei $^{1}$ \\
 $^1$Math. Dept., College of Science, University of
Bahrain, Kingdom of Bahrain\\
$^2$ Math. Dept., Faculty of science,  Aswan University, Sahari
81528, Aswan, Egypt
\end{center}

\begin{abstract}
Some properties of multi-qubit systems interacting with noisy
environment is discussed. The amount of the survival entanglement
is quantified for the GHZ and W-states. It is shown that the
entanglement decay  depends on the noise type ( correlated or
non-correlated), number of interacted qubits with the environment
and the initial state which passes through this noisy environment.
In general, the  GHZ is more fragile than the  W-state. The
phenomena of entanglement sudden death appears only for
non-correlated noise.

\end{abstract}

\section{Introduction}

Entanglement for  two-qubit systems has been investigated
extensively in many directions. Quantifying  and manipulating
entanglement for  systems of small dimension  have been covered a
in wide area of research \cite{Peres, Horodecki,Zyc}. However,
there are a lot of applications  that have been implemented  on
quantum information and computations by using these systems
theoretically and experientially. Quantum teleportation, coding
\cite{metwally1} and cryptography represent the more important
applications of entanglement. To implement these application with
high efficiency, maximum entangled states (MES) are needed.
Despite  it is possible to generate maximum entangled states, they
turn into partially entangled states (PES) due to their
interaction with their own environments \cite{Nielsen}. Therefore,
another process is needed to recover entanglement, called quantum
purification, which requires infinite number of pairs to get one
MES \cite{bennt}. So, it is important to investigate the behavior
of entanglement in the presences of different types of noise.
There are some classes of noise states whose teleportation
fidelity can be enhanced if one of the two qubits subject to
dissipative interaction with the environment via amplitude damping
channel \cite{Badziag2002}.

 Recently, Montealegre et. al.
\cite{Mont2013}, have investigated the effect of  different types
of noise channels on the one-norm geometric discord and showed
that the effect of the noise channel can be frozen. Metwally
showed that the  phenomena of sudden single and  double  changes
and the sudden death of entanglement are reported for identical
and non-identical noise \cite{metwally3}.  Also, the behavior of a
multi-qubit system in the presence of the  amplitude damping
channel is investigated in \cite{metwally4}, where the possibility
of performing quantum teleportation with decohered states was
discussed. This motivates us to investigate two classes of
multi-qubit systems, the GHZ and  the W-states, where  the effect
of the correlated or non-correlated noisy depolarized channel is
considered.

The paper is organized as  follows: In Sec. $2$, we describe the
initial state and their evolution in the presence of depolarized
damping channel. The behavior of entanglement for the GHZ state is
discussed in Sec. $3$, while  Sec. $4$, is devoted for the
W-state. Finally we conclude our results in Sec. $5$.

\section{The system}
Assume that  a source generates a three-qubit state in the form of
GHZ or W-state defined as:

\begin{eqnarray}\label{iniS}
 \ket{\psi_{ini}}&=& \left\{ \begin{array}{ll}
\ket{\psi_{g}}=\frac{1}{\sqrt{2}}(\ket{000}+\ket{111}),&  \\
\ket{\psi_{w}}=\frac{1}{\sqrt{3}}(\ket{100}+\ket{010}+\ket{001}).& \\
\end{array} \right.
\end{eqnarray}
 During the transmission from the source to the users, Alice, Bob, and Charlie,
  it is assumed that Alice's and Bob's qubits passes through
 the  depolarized channel \cite{Nielsen}. This channel transforms
 the single qubit  into a completely mixed state with probability
 $p$ and leaves it untouched  with probability $1-p$. For a single
 qubit noise, the operators are given by

\begin{eqnarray}
\mathcal{D}^{(j)}_1&=&\sqrt{1-p}~I^{(j)}, \quad
\mathcal{D}^{(j)}_2=\sqrt{\frac{p}{3}}\sigma_x^{(j)},\quad
\mathcal{D}^{(j)}_3=\sqrt{\frac{p}{3}}\sigma_y^{(j)},\quad
\mathcal{D}^{(j)}_4=\sqrt{\frac{p}{3}}\sigma_z^{(j)},
\end{eqnarray}
where $I^{(j)}_i,\sigma^{(j)}_i, i=x,y,z, j=a,b$ and $c$ are the
identity  and  Pauli operators for the three qubits respectively.
In  computational basis, these operators can be written as:
\begin{eqnarray}
I^{(j)}_i&=&\ket{0}\bra{0}+\ket{1}\bra{1},\quad
\sigma_x^{(j)}=\ket{0}\bra{1}+\ket{1}\bra{0},\quad
 \nonumber\\
 \sigma_y^{(j)}&=&i(\ket{0}\bra{1}-\ket{1}\bra{1})\quad
\sigma_z^{(j)}=\ket{0}\bra{0}-\ket{1}\bra{0}.
\end{eqnarray}
We  consider  that the suggested noise model is of   one, two and
three  sides noisy channel type. For two and three sides noise, we
consider that the noise could be correlated or non-correlated
\cite{Nagwa}. For correlated noise, we mean that the qubit is
affected by the same noise at the same time, while for
non-correlated noise, the qubits may  be affected  by different
noises at the same time.

The three-qubit system under the depolarizing noise acting on the
first qubit of the quantum state $\rho_{ini}$ is given by
\begin{equation}
\rho^{(f_i)}=\sum_{k=1}^{4}\Bigl\{\mathcal{D}^{a}_k\otimes
I^{(b)}\otimes I^{(c)}\rho_{ini}I^{(c)}\otimes I^{(b)}\otimes
\mathcal{D}_k^{\dagger(a)}\Bigl\},
\end{equation}
where $\rho_{ini}=\ket{\psi}_{ini}\bra{\psi}$ and
$\ket{\psi_{ini}}$ is one of the initial states (1). However, if
the first two qubits are affected by the depolarized channel, then
the initial state $\rho_{ini}$  evolves as
\begin{equation}
\rho^{(f_{co_2})}=\sum_{k=1}^{4}\Bigl\{\mathcal{D}^{(a)}_k\otimes
\mathcal{D}_k^{(b)}\otimes I^{(c)}\rho_{ini}I^{(c)}\otimes
\mathcal{D}_k^{\dagger(b)}\otimes
\mathcal{D}_k^{\dagger(a)}\Bigl\},
\end{equation}
for correlated noise, while for non-correlated noise, the final
state can be written as $\rho^{(f)}$
\begin{equation}
\rho^{(f_{nc_2})}=\sum_{k=1}^{k=4}\sum_{\ell=1}^{\ell=4}\Bigl\{\mathcal{D}^{(a)}_k\otimes
\mathcal{D}_{\ell}^{(b)}\otimes I^{(c)}\rho_{ini}I^{(c)}\otimes
\mathcal{D}_{\ell}^{\dagger(b)}\otimes
\mathcal{D}_k^{\dagger(a)}\Bigl\}.
\end{equation}
Finally, if the three qubits are affect by the depolarizing
channel (2), then for correlated noise, the final state of the
initial three qubits (1) evolves as,

\begin{equation}
\rho^{(f_{co_3})}=\sum_{k=1}^{k=4}\Bigl\{\mathcal{D}^{(a)}_k\otimes
\mathcal{D}_k^{(b)}\otimes
\mathcal{D}_k^{(c)}\rho_{ini}\mathcal{D}_k^{\dagger(c)}\otimes
\mathcal{D}_k^{\dagger(b)}\otimes
\mathcal{D}_k^{\dagger(a)}\Bigl\},
\end{equation}
while for the non-correlated noise it is given by

\begin{equation}
\rho^{(f_{nc_3})}=\sum_{k=1}^{k=4}\sum_{\ell=1}^{\ell=4}\sum_{m=1}^{m=4}\Bigl\{\mathcal{D}^{(a)}_k\otimes
\mathcal{D}_{\ell}^{(b)}\otimes
\mathcal{D}_k^{(c)}\rho_{ini}\mathcal{D}_k^{\dagger(c)}\otimes
\mathcal{D}_{\ell}^{\dagger(b)}\otimes
\mathcal{D}_k^{\dagger(a)}\Bigl\},
\end{equation}
where $\rho^{(f_{co_r})}$ and  $\rho^{(f_{nc_r})}$ when $r=2,3$
stands for the final states for correlated and non-correlated
noise if $2$ or $3$ qubits are affected, respectively.

In the following we shall investigate the dynamics of entanglement
which is contained in the initial states (1), where we investigate
all the possible noises as described above. For this aim, we  use
 the tripartite negativity as a measure of entanglement. This
measure states that, if $\rho_{abc}$ represents a tripartite
state, then the negativity is defined as \cite{Carlos},
\begin{equation}
\mathcal{N}(\rho_{abc})=(\mathcal{N}_{a-bc}\mathcal{N}_{b-ac}\mathcal{N}_{c-ab})^{\frac{1}{3}},
\end{equation}
where
$\mathcal{N}_{i-jk}=-2\sum_{\ell}{\lambda_{\ell}(\rho_{ijk}^{T_i})}$,
$\lambda_{\ell}$ are the negative eigenvalues of the partial
transpose of the state $\rho_{ijk}$ with respect to the qubit
$"i"$
\section{GHZ-state}
\begin{enumerate}
\item{Correlated Noise:\\}
 The initial three-qubit state
$\rho_g=\ket{\psi_g}\bra{\psi_g}$, $\ket{\psi_g}$ is given from
(1) under the depolarizing noise on Alice's qubit is given by
\begin{eqnarray}
\rho^{(f)}_{g}&=&(1-p)\rho_{g}+
\frac{p}{3}\Bigl(\ket{111}\bra{100}- \ket{111}\bra{011}\Bigr)
\nonumber\\
&&+\frac{p}{6}\Bigl(\ket{011}\bra{100}+\ket{100}\bra{011}
\ket{100}\bra{100}+ \ket{011}\bra{011}\Bigr).
\end{eqnarray}
However, if we assume that the first two qubits subject to
correlated noise then the final state (5) is given by

\begin{equation}
\rho_g^{(f_{co_2})}=\Bigl(\frac{1}{2}(1-p)^2+\frac{p^2}{18}\Bigr)\rho_g+
\frac{p^2}{9}\Bigl(\ket{001}\bra{110}+\ket{110}\bra{001}+\ket{001}\bra{001}+\ket{110}\bra{110}\Bigr).
\end{equation}
Finally, if it is assumed that all the three qubits are subject to
a correlated noise, then $\rho^{(f_{co_3})}$ can be written as

\begin{figure}[t!]
  \begin{center}
  \includegraphics[width=25pc,height=15pc]{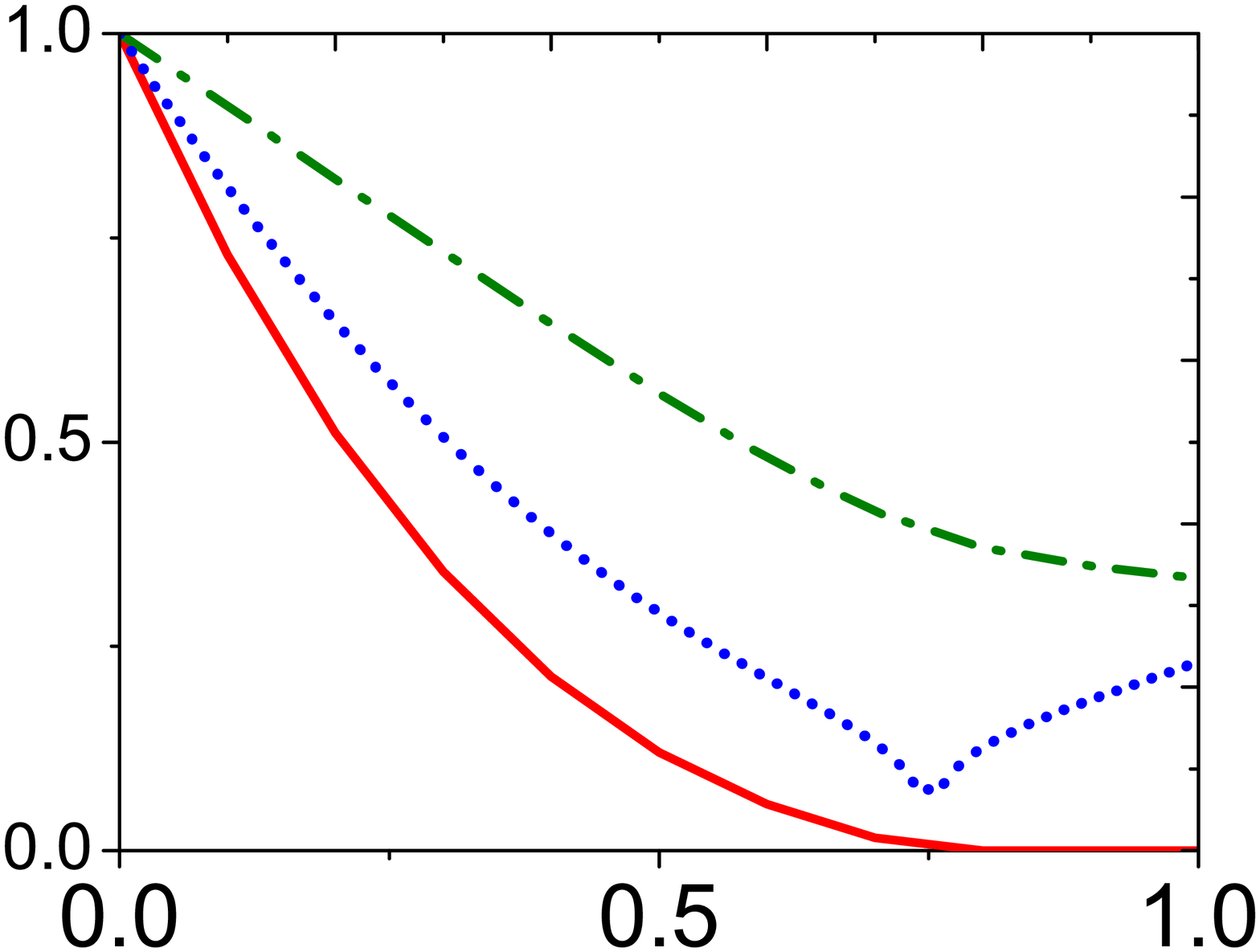}~\quad
      \put(-310,90){$\mathcal{N}(\rho^{(f)})$}
 \put(-145,-1){\Large$p$}
     \caption{The entanglement $\mathcal{N}(\rho_g^{(f)})$ of the GHZ state  under depolarize channel. The
     solid and the dot curves for the  correlated and non-correlated noise respectively noise.}
       \end{center}
\end{figure}

\begin{eqnarray}
\rho_g^{(f_{co_3})}&=&\Bigl(\frac{(1-p)^3}{2}+\frac{p^2}{18}\Bigr)\ket{000}\bra{000}+
\Bigl(\frac{(1-p)^3}{2}-\frac{p^3}{54}\Bigr)\ket{000}\bra{111}
\nonumber\\
&&+\frac{(1-p)^3}{2}\ket{111}\bra{000}+\Bigl(\frac{(1-p)^3}{2}+\frac{p^2}{27}\Bigr)\ket{111}\bra{111}.
\end{eqnarray}
In Fig.(1), we investigate the behavior of entanglement for the
three final states $(9-11)$. The general behavior displays that
the entanglement decays as $p$ increases. However, these decays
depends on the number of the affected qubits with the depolarized
channel. For the state $\rho_g^{(f)}$, where only one qubit is
depolarized, the entanglement decays smoothly, while  for $p=1$,
when  the strength of noise is maximum, the entanglement doesn't
vanish. For the  state $\rho_g^{(f_{co_2})}$, where two qubits are
depolarized,  the  decay rate  is larger than that depicted for
the state $\rho_g^{(f)}$. Although the entanglement vanishes for
some specific value of $p\simeq 0.75$, it re-increases again  for
farther values of $p$. Finally,  the  decay rate of entanglement
is the largest one if the three qubits are depolarized as shown
for the state  $\rho_g^{(f_{co_3})}$.  For this state the
entanglement decays gradually until vanishes completely at
$p\simeq 0.78$.

\item{Non-correlated noise:\\} Now, we assume that two or three
qubits are depolarized in non-correlated way. If only two particle
are depolarized then the final sate (6) can be written as

\begin{eqnarray}
\rho_g^{(f_{nco_2})}&=&\mathcal{A}_1\Bigl(\ket{000}\bra{000}+\ket{111}\bra{111}\Bigr)+
\mathcal{A}_2\Bigl(\ket{000}\bra{111}+\ket{111}\bra{000}\Bigr)
\nonumber\\
&&+\mathcal{A}_3\Bigl(\ket{010}\bra{010}+\ket{101}\bra{101}+\ket{011}\bra{011}+\ket{100}\bra{100}\Bigr)
\nonumber\\
&&+\mathcal{A}_4\ket{001}\bra{110}+\mathcal{A}_5\ket{001}\bra{001}+\mathcal{A}_6\ket{110}\bra{110}+
\mathcal{A}_7\ket{110}\bra{001}
\nonumber\\
&&+\mathcal{A}_8\ket{001}\bra{111}+\mathcal{A}_9\Bigl(\ket{001}\bra{011}+\ket{110}\bra{100}-\ket{110}\bra{011}
\Bigr),
\end{eqnarray}
where
\begin{eqnarray}
\mathcal{A}_1&=&
\frac{(1-p)^2}{2}+\frac{p}{3}(1-p)+\frac{p^2}{18}, \quad
\mathcal{A}_2=\frac{(1-p)^2}{2}-\frac{p}{3}(1-p)+\frac{p^2}{18},
\nonumber\\
\mathcal{A}_3&=&\frac{p}{3}(1-p)+\frac{p^2}{9}, \quad
\mathcal{A}_4=\frac{p^2}{9}-\frac{p}{3}(1-p), \quad
\mathcal{A}_5=\frac{p^2}{18},
\mathcal{A}_6=\frac{p^2}{9}+\frac{p}{6}(1-p),
\nonumber\\
\mathcal{A}_7&=&\frac{p^2}{9}-\frac{p}{6}(1-p), \quad
\mathcal{A}_8=\frac{p}{6}(1-p), \quad
\mathcal{A}_9=\frac{p^2}{18}.
\end{eqnarray}

\begin{figure}[t!]
  \begin{center}
  \includegraphics[width=25pc,height=15pc]{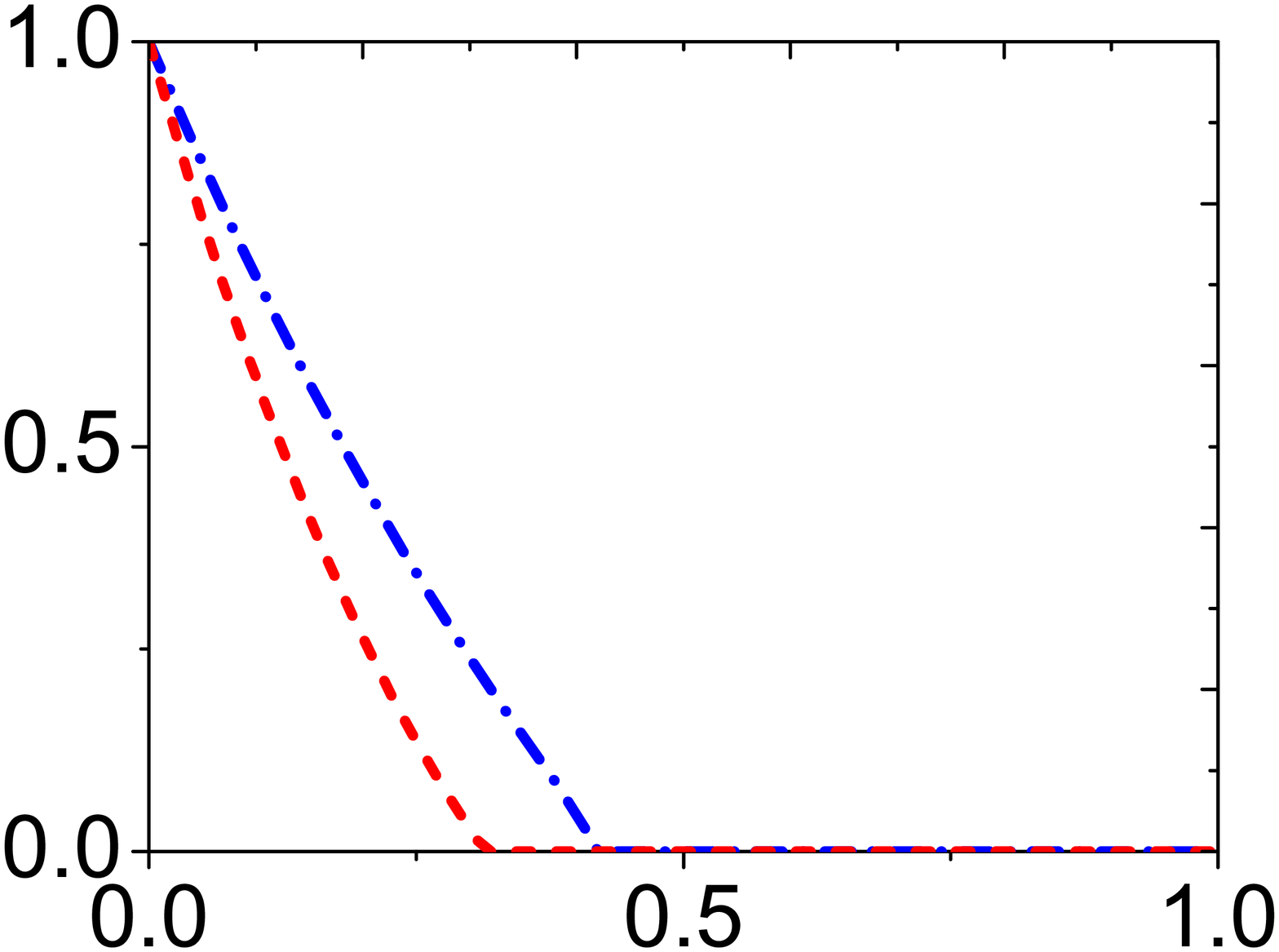}\quad
      \put(-310,90){$\mathcal{N}(\rho^{(f)})$}
 \put(-145,-1){\Large$p$}
     \caption{The dash-dot curves represents the entanglement $\mathcal{N}(\rho^{(f)})$, where it is assumed
     only two qubits are depolarized, while the dot curve when the 3-qubits are depolarized.  }
       \end{center}
\end{figure}

However, if we assume that the three qubits pass through the
depolarized channel (2), then the final state (8) (for
non-correlated noise) is given by

\begin{eqnarray}
\rho_g^{(f_{nco_3})}&=&\mathcal{B}_1\left(\ket{000}\bra{000}+\ket{111}\bra{111}\right)+
\mathcal{B}_2\left(\ket{000}\bra{111}+\ket{111}\bra{000}\right)
\nonumber\\
&&+\mathcal{B}_3\left(\ket{110}\bra{110}+\ket{001}\bra{001}+\ket{100}\bra{100}+\ket{011}\bra{011}\right)
\nonumber\\
&&+\mathcal{B}_4\left(\ket{110}\bra{001}+\ket{010}\bra{101}+\ket{101}\bra{010}+\ket{100}\bra{011}+\ket{011}\bra{100}\right)
\nonumber\\
&&+\mathcal{B}_5\left(\ket{001}\bra{110}+\mathcal{B}_6(\ket{010}\bra{010}+\ket{101}\bra{101}\right),
\end{eqnarray}

where
\begin{eqnarray}
\mathcal{B}_1&=&\frac{1}{2}\left[(1-p)^3+p(1-p)+\frac{5}{9}p^2(1-p)+\frac{p^3}{3}\right],
\nonumber\\
\mathcal{B}_1&=&\frac{1}{2}\left[(1-p)^3-p(1-p)+\frac{5}{9}p^2(1-p)-\frac{p^3}{27}\right],
\nonumber\\
\mathcal{B}_3&=&\frac{1}{3}\left[p(1-p)^2+\frac{7p^2}{3}(1-p)+\frac{p^3}{3}\right],
\nonumber\\
\mathcal{B}_4&=&\frac{p^2}{9}(1-p),\quad
\mathcal{B}_5=\frac{p}{2}(1-p)^2+\frac{p^2}{9}(1-p),
\nonumber\\
\mathcal{B}_6&=&\frac{1}{3}\left[p(1-p)^2+\frac{5p^2}{3}(1-p)+\frac{p^3}{3}\right].
\end{eqnarray}

Fig.(2), shows the behavior of entanglement which is contained in
the final sates $\rho_g^{(f_{nco_3})}$ and $\rho_g^{(f_{nco_3})}$,
where it is assumed that the noise is non-correlated. The general
behavior shows that the entanglement decays hastily as the channel
parameter $p$ increases. The decay rate of entanglement depends on
the numbers of depolarized qubits. However, the decay rate of
entanglement when the three qubit subject to the noise is larger
than that depicted when only two qubits are depolarized. The
phenomena of the sudden-death of entanglement appear clearly when
the particles are affected by non-correlated noise.

From Fig.(1) and Fig.(2), we can conclude that the decay of
entanglement depends on the type of the noise (correlated or
non-correlated) and the number of qubits which  are subject to the
noise. For correlated noise, the entanglement decays smoothly to
reach its minimum bounds. These bounds depend on the number of
polarized qubits, where  they are non-zero for fewer number of
polarized particles and completely vanish as the number of
depolarized qubit increases. However, the phenomena of the sudden
death appears in the presences of non-correlated noise and as the
number of polarized qubit increases, the entanglement dies for
smaller values of the channel parameter $p$.

\end{enumerate}

\section{W-state}

Now we assume that the user share a three-qubits-state of W-type
$\rho_w=\ket{\psi_w}\bra{\psi_w}$, where $\ket{\psi_w}$ is defined
in (1). If we allow only one qubit to pass through the depolarized
channel (3), then the final state is given by
\begin{eqnarray}
\rho^{(f)}_w&=&\frac{3-2p}{9}\Bigl\{\ket{100}\bra{100}+\ket{010}\bra{010}+\ket{001}\bra{010}+\ket{001}\bra{001}
+\ket{010}\bra{001}\Bigr\}
\nonumber\\
&&+\frac{3-4p}{9}\Bigl\{\ket{100}\bra{010}+\ket{100}\bra{001}+\ket{010}\bra{100}+\ket{001}\bra{100}\Bigr\}.
\end{eqnarray}
However, if we assume that only two qubits are forced to pass
through  the noise channel (3) and that the noise is correlated,
then the final state is given by,
\begin{eqnarray}
\rho_w^{f_{cor_2}}&=&
\tilde\mathcal{A}_1\Bigl\{\ket{010}\bra{010}+\ket{010}\bra{100}+\ket{100}\bra{010}+\ket{100}\bra{100}\Bigr\}
\nonumber\\
&&+\tilde\mathcal{A}_2\Bigl\{\ket{010}\bra{001}+\ket{001}\bra{100}+\ket{001}\bra{010}\Bigr\}
\nonumber\\
&&+\tilde\mathcal{A}_3\Bigl\{\ket{100}\bra{011}+\ket{001}\bra{001}\Bigr\}+\tilde\mathcal{A}_4\ket{010}\bra{101}+
\tilde\mathcal{A}_5\ket{111}\bra{111},
\end{eqnarray}
where
\begin{eqnarray}
\tilde\mathcal{A}_1&=&\frac{(1-p)^2}{3}+\frac{P^2}{9},\quad
\tilde\mathcal{A}_2=\frac{(1-p)^2}{3}-\frac{P^2}{27} \nonumber\\
\tilde\mathcal{A}_3&=&\frac{(1-p)^2}{3}+\frac{P^2}{27},\quad
\tilde\mathcal{A}_4=\frac{p^2}{27}, \quad
\tilde\mathcal{A}_5=\frac{2p^2}{27}
\end{eqnarray}

\begin{figure}[t!]
  \begin{center}
  \includegraphics[width=25pc,height=15pc]{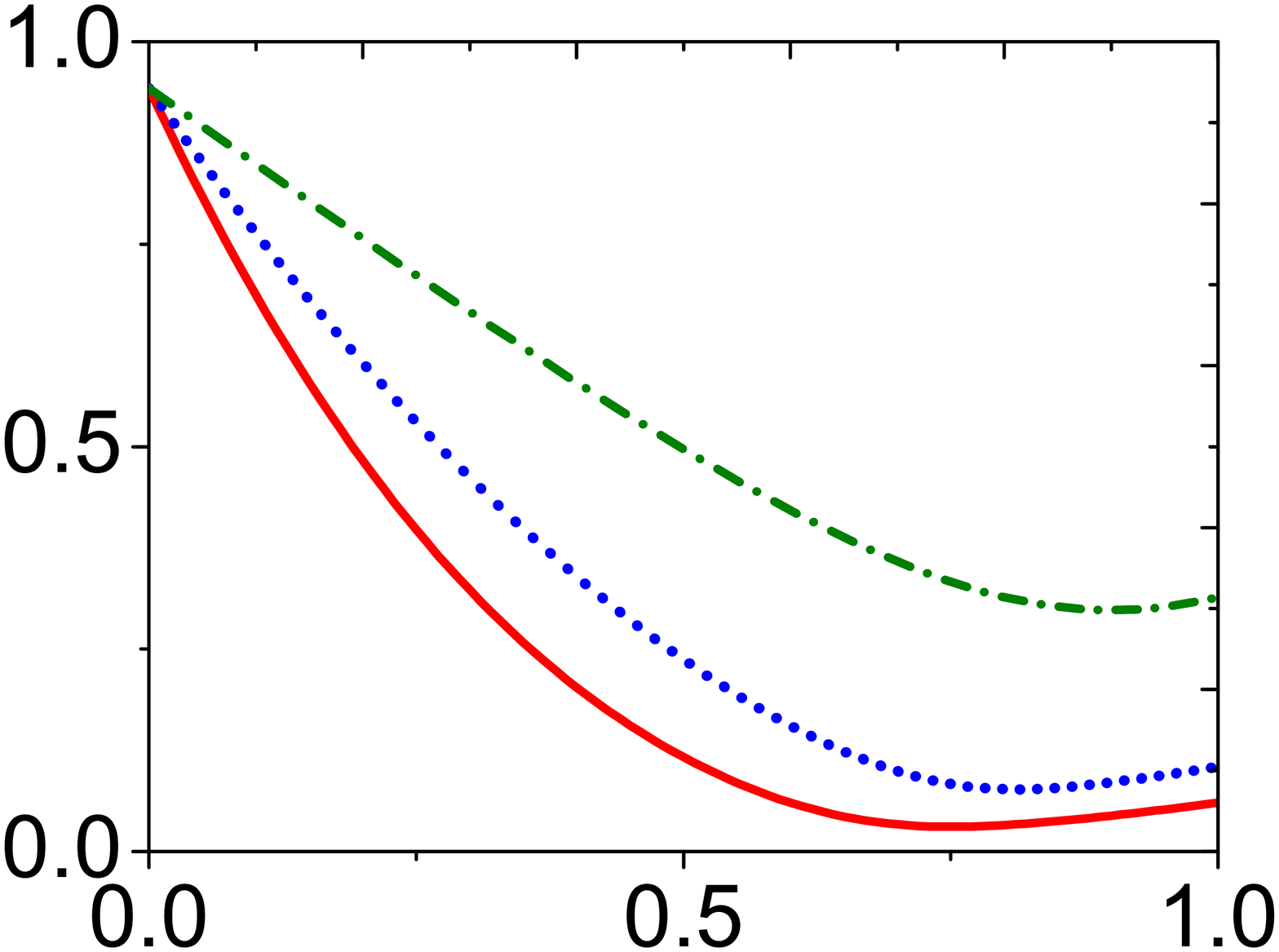}\quad
      \put(-310,90){$\mathcal{N}(\rho^{(f)})$}
 \put(-145,-1){\Large$p$}
     \caption{The dash-dot, dot and solid   curves represent the
     entanglement of the final $\rho_w^{(f)}$
     when one, two and three qubits pass through the depolarized channel(3)in  a correlated way, respectively. }
       \end{center}
\end{figure}
However, if the three qubits pass through the non-correlated noise
then the final state is given by
\begin{eqnarray}
\rho^{(f_{co_3)}}_w&=&\tilde\mathcal{B}_1\Bigl\{\ket{001}\bra{001}+\ket{001}\bra{010}+\ket{001}\bra{100}\Bigr\}
+\tilde\mathcal{B}_2\ket{010}\bra{001}+\tilde\mathcal{B}_3\ket{100}\bra{011}
\nonumber\\
&&+\tilde\mathcal{B}_4\Bigl\{\ket{100}\bra{100}+\ket{100}\bra{010}+\ket{010}\bra{010}+\ket{010}\bra{100}\Bigr\}
\nonumber\\
&&+\tilde\mathcal{B}_5\Bigl\{\ket{010}\bra{101}+\ket{010}\bra{111}+\ket{010}\bra{011}-\ket{100}\bra{111}
-\ket{111}\bra{010}
\nonumber\\
&&\hspace{7cm}-\ket{111}\bra{100}-\ket{111}\bra{111}\Bigr\},
\end{eqnarray}
where
\begin{eqnarray}
\tilde\mathcal{B}_1&=&\frac{(1-p)^2}{3}+\frac{2p^2}{27}, \quad
\tilde\mathcal{B}_2=\frac{(1-p)^2}{3}-\frac{p^2}{27},
 \nonumber\\
 \quad
 \tilde\mathcal{B}_3&=&\frac{(1-p)^2}{3}+\frac{p^2}{27},\quad
 \tilde\mathcal{B}_4=\frac{(1-p)^2}{3}+\frac{2p^2}{9},
\quad \tilde\mathcal{B}_5=\frac{p^2}{27}.
\end{eqnarray}

Fig.(3), describes the behavior of entanglement for the final
states (16),(17) and (19), where we assume that, one, two and
three qubits are subjected to the noise channel (3), respectively.
The general behavior shows that the entanglement decays as the
channel parameter $p$ increases. The decay rate depends on the
number of depolarized qubits. It is clear that, if only one qubit
is depolarized, then the entanglement decays smoothly to reach its
minimum non-zero value. These minimum values decrease as the
number of depolarized qubits increases.
\begin{figure}
  \begin{center}
  \includegraphics[width=25pc,height=15pc]{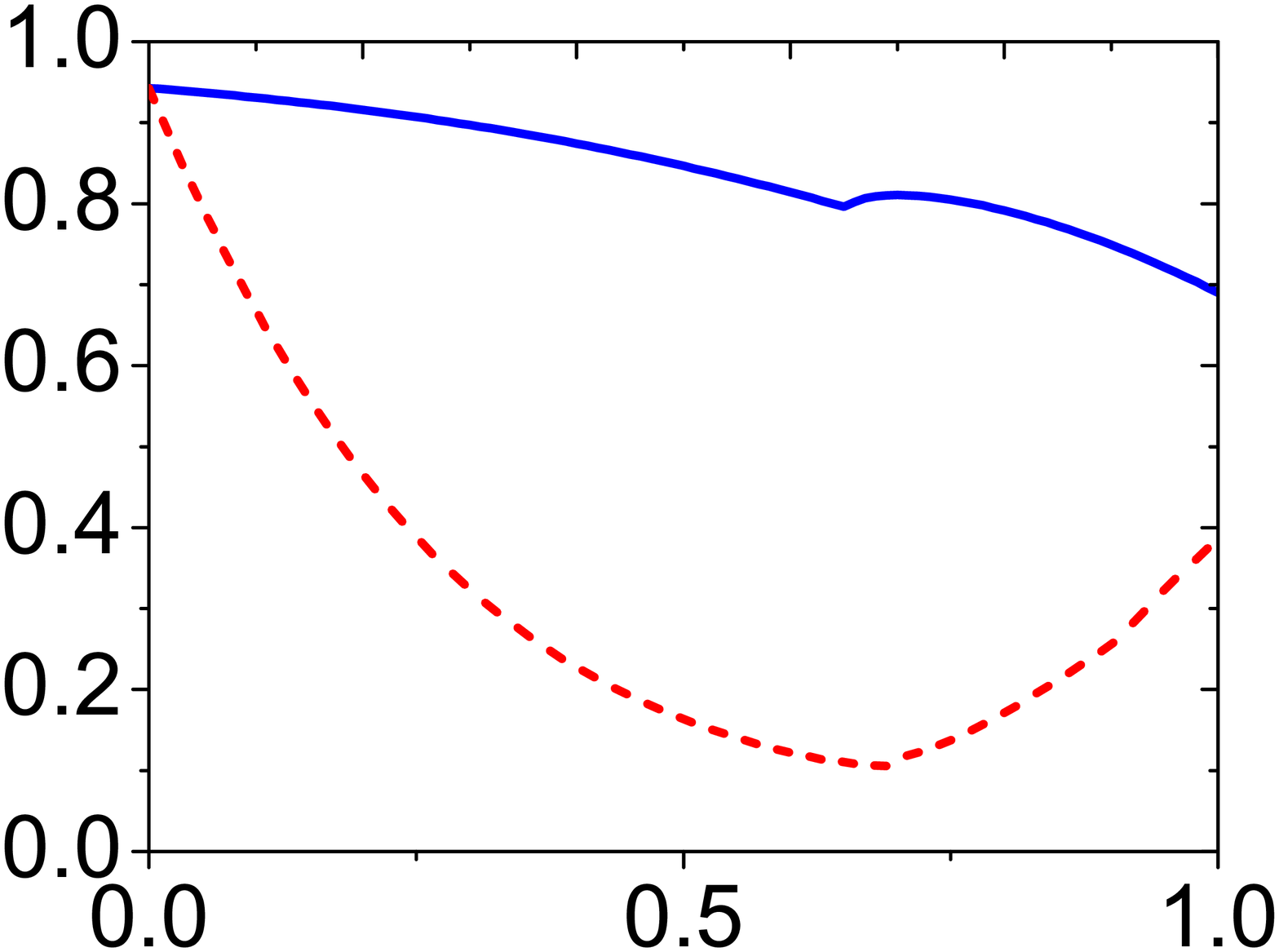}\quad
      \put(-330,90){$\mathcal{N}(\rho_w^{(f_{co2})})$}
 \put(-145,-1){\Large$p$}
     \caption{The  solid and dash curves represent the
     entanglement of the final states  $\rho_w^{(f_{co_2})}$ and $\rho_w^{(f_{co_3})}$, respectively. }
       \end{center}
\end{figure}

Now, we assume that the effect of the is non-correlated. If only
two  qubits  are passing through the depolarized channel, then the
noise output state is given by,
\begin{eqnarray}
\rho_w^{f_{nc_2}}&=&\tilde\mathcal{A}_1^{nc}\Bigl\{\ket{000}\bra{011}+\ket{000}\bra{101}+\ket{011}\bra{00}+\ket{101}\bra{000}\Bigr\}
\nonumber\\
&&+\tilde\mathcal{A}_2^{nc}\Big\{\ket{010}\bra{010}+\ket{100}\bra{100}+\ket{001}\bra{001}\Bigl\}+\tilde\mathcal{A}_3^{nc}\ket{010}\bra{100}
\nonumber\\
&&+\tilde\mathcal{A}_3^{nc}\Bigl\{\ker{001}\bra{010}+\ket{001}\bra{100}+\ket{100}\bra{001}+\ket{010}\bra{001}\Bigl\}
\nonumber\\
&&+\tilde\mathcal{A}_4^{nc}\Bigl\{\ket{110}\bra{000}+\ket{000}\bra{110}\Bigr\}+
\tilde\mathcal{A}_5^{nc}\ket{100}\bra{010},
\end{eqnarray}
where
\begin{eqnarray}
\tilde\mathcal{A}_1^{nc}&=&\frac{2p}{27},\quad
\tilde\mathcal{A}_2^{nc}=\frac{2}{27}p(1-p)+\frac{p^2}{27}+\frac{1}{3},
\quad
\tilde\mathcal{A}_3^{nc}=-\frac{2}{27}p(1-p)-\frac{5p^2}{27}+\frac{1}{3},
\nonumber\\
\tilde\mathcal{A}_4^{nc}&=&\frac{4p}{27}, \quad
\tilde\mathcal{A}_5^{nc}=\frac{p^2}{27}+\frac{1}{3}, \quad
\tilde\mathcal{A}_6^{nc}=\frac{2}{27}\bigl\{3p^2-p\bigr\}+\frac{1}{3}.
\end{eqnarray}
However, we assume that the three qubits of the  W-states passing
through the noise channel. In this case the output state is given
by
\begin{eqnarray}
\rho_w^{(f_{nc_3})}&=&\tilde\mathcal{B}_1^{nc}\Bigl\{\ket{110}\bra{000}+\ket{000}\bra{110}+\ket{101}\bra{000}\Bigr\}
+\tilde\mathcal{B}_2^{nc}\Big\{\ket{011}\bra{000}+\ket{000}\bra{101}\Bigr\}
\nonumber\\
&&+\tilde\mathcal{B}_3^{nc}\ket{111}\bra{100}+\tilde\mathcal{B}_4^{nc}\ket{011}\bra{011}+\tilde\mathcal{B}_5^{nc}\ket{101}\bra{011}
+\tilde\mathcal{B}_6^{nc}\ket{000}\bra{011}+\tilde\mathcal{B}_7^{nc}\ket{010}\bra{010}
\nonumber\\
&&+\tilde\mathcal{B}_8^{nc}\Bigl\{\ket{111}\bra{111}+\ket{110}\bra{011}+\ket{011}\bra{100}+\ket{011}\bra{111}+\ket{101}\bra{101}\Bigl\}
\nonumber\\
&&+
\tilde\mathcal{B}_{9}^{nc}\Bigl\{\ket{001}\bra{010}+\ket{100}\bra{010}+\ket{010}\bra{001}+\ket{010}\bra{100}\Bigr\}
\nonumber\\
 &&+\tilde\mathcal{B}_{10}^{nc}\Bigl\{\ket{100}\bra{001}+\ket{001}\bra{100}\Bigr\}+\tilde\mathcal{B}_{11}^{nc}\Bigl\{\ket{001}\bra{001}+\ket{100}\bra{100}\Bigr\}
\nonumber\\
&&+\tilde\mathcal{B}_{12}^{nc}\Bigl\{\ket{011}\bra{010}+\ket{111}\bra{010}+
\ket{110}\bra{101} \Bigr\},
 \end{eqnarray}
where
\begin{eqnarray}
\tilde\mathcal{B}_1^{nc}&=&\frac{4}{3^2}p(1-p)^2+\frac{1}{3^2}p^2(1-p)-\frac{4}{3^4}p^3,\quad
\tilde\mathcal{B}_2^{nc}=\frac{4}{3^2}p(1-p)^2-\frac{4}{3^4}p^3,
\nonumber\\
\tilde\mathcal{B}_3^{nc}&=&
\frac{1}{3^2}p^2(1-p)+\frac{1}{3^4}p^3,\quad
\tilde\mathcal{B}_4^{nc}=\frac{2}{3^4}p^3,\quad
\tilde\mathcal{B}_5^{nc}=-\frac{1}{3^3}p^2(1-p)-\frac{2}{3^4}p^3,
\nonumber\\
\tilde\mathcal{B}_6^{nc}&=&\frac{4}{3^2}p(1-p)^2-\frac{2}{3^4}p^3,
\quad
\tilde\mathcal{B}_7^{nc}=(1-p)^3+\frac{4}{3^2}p(1-p)^2+\frac{1}{3^2}p^2(1-p)+\frac{1}{3^4}p^3,
\nonumber\\
\tilde\mathcal{B}_8^{nc}&=&\frac{1}{3^3}p^2(1-p),\quad
\tilde\mathcal{B}_{9}^{nc}=\frac{1}{3}(1-p)^3-\frac{2}{3^2}p(1-p)^2+\frac{1}{3^2}p^2(1-p)+\frac{5}{3^4}p^3,
\nonumber\\
\tilde\mathcal{B}_{10}^{nc}&=&\frac{(1-p)^3}{3}+\frac{p^2}{3^2}(1-p)+\frac{5p^3}{3^4},\quad
\tilde\mathcal{B}_{11}^{nc}=\frac{(1-p)^3}{3}+\frac{2p}{3^2}(1-p)^2+\frac{p^2}{3^2}(1-p)+\frac{p^3}{3^4},
\nonumber\\
\tilde\mathcal{B}_{12}^{nc}&=&-\frac{p^2}{3^3}(1-p).
\end{eqnarray}

The behavior of entanglement for the output states (21) and (23)
is described in Fig.(4).
 In general, the entanglement decreases as the channel's strength increases. However the
decreasing rate  is smaller than that depicted for correlated
noise (see Fig.3). Moreover, when  all three qubits   pass through
the depolarized channel, the entanglement decays faster than that
depicted for two depolarized qubits.  It is clear that for larger
values of the strength parameter $p$, the entanglement increases.
However, the upper bound of entanglement is smaller than  that in
the case of two-depolarized qubit.

 On the other hand, comparing the behavior of
this entanglement with that for the  GHZ state, we can see that
the W-state is more robust than the  GHZ state, where for the GHZ
state the entanglement  completely vanishes in a  sudden way  for
smaller values of the channel strength.

\section{Conclusion}
In this contribution, we investigate the behavior of entanglement
for two classes of multi-qubit states; the GHZ and W-states
passing trough the depolarized channel. In our treatment, we
assume that the effect of the noise channel includes two
possibilities. In the first possibility we assume that the qubits
are subject to the same noise at the same time (correlated noise).
The second possibility,  which is called non-correlated noise,
 the particles are subjected to different noises at the same
time. Analytical expressions are obtained for the final output
states. For the correlated noise, we calculate the final output
states when one, two or  three qubits  are forced to pass through
the depolarized channel. However, for the non-correlated noise, we
consider the case of two or three qubits passing through the
depolarized channel.

The dynamics of the survival amount of entanglement is
investigated against  the channel strength. The general behavior
shows that the entanglement  decreases as the channel parameter
increases. However, the decreasing rate depends on the number of
depolarized qubits, the type of the noise and the type of the
initial state. For {\it correlated} noise, the decay rate of
entanglement decreases as the number of polarized qubits
increases. Although the decay rate for the  W-state is larger than
that depicted for the  GHZ state, the entanglement  completely
vanishes when  the three qubits are depolarized for the  GHZ
state. On the other hand, for the {\it non-correlated} noise, the
phenomena of the sudden death of entanglement appears when the
GHZ's qubits  are depolarized. If only two qubit qubits are
depolarized, the entanglement vanishes at larger values of the
channel parameter. However, for the  W-state, the phenomena of the
entanglement sudden death doesn't appear and the lower bound of
entanglement is large.

{\it In conclusion}, the robustness or fragility of the initial
states which pass through a depolarized channel depends on the
type of the noise: correlated or non-correlated. In general GHZ is
more fragile than W-state. The phenomena of entanglement sudden
death appears only for non-correlated noise.

\end{document}